\documentclass[twocolumn,showpacs,preprintnumbers,amsmath,amssymb]{revtex4}

\usepackage{graphicx}
\usepackage{dcolumn}
\usepackage{bm}

\begin{document}
\newcommand{\Arg}[1]{\mbox{Arg}\left[#1\right]}
\newcommand{\bb}{\mathbf}
\newcommand{\braopket}[3]{\left \langle #1\right| \hat #2 \left|#3 \right \rangle}
\newcommand{\braket}[2]{\langle #1|#2\rangle}
\newcommand{\be}{\[}
\newcommand{\br}{\vspace{4mm}}
\newcommand{\bra}[1]{\langle #1|}
\newcommand{\braketbraket}[4]{\langle #1|#2\rangle\langle #3|#4\rangle}
\newcommand{\braop}[2]{\langle #1| \hat #2}
\newcommand{\dd}[1]{ \! \! \!  \mbox{d}#1\ }
\newcommand{\DD}[2]{\frac{\! \! \! \mbox d}{\mbox d #1}#2}
\renewcommand{\det}[1]{\mbox{det}\left(#1\right)}
\newcommand{\ee}{\]} 
\newcommand{\eg}{\textbf{\\  Example: \ \ \ }}
\newcommand{\Imag}[1]{\mbox{Im}\left(#1\right)}
\newcommand{\ket}[1]{|#1\rangle}
\newcommand{\ketbra}[2]{|#1\rangle \langle #2|}
\newcommand{\kp}{\arccos(\frac{\omega - \epsilon}{2t})}
\newcommand{\ldos}{\mbox{L.D.O.S.}}
\renewcommand{\log}[1]{\mbox{log}\left(#1\right)}
\newcommand{\Log}{\mbox{log}}
\newcommand{\Modsq}[1]{\left| #1\right|^2}
\newcommand{\nb}{\textbf{Note: \ \ \ }}
\newcommand{\op}[1]{\hat {#1}}
\newcommand{\opket}[2]{\hat #1 | #2 \rangle}
\newcommand{\occ}{\mbox{Occ. Num.}}
\newcommand{\Real}[1]{\mbox{Re}\left(#1\right)}
\newcommand{\so}{\Rightarrow}
\newcommand{\sol}{\textbf{Solution: \ \ \ }}
\newcommand{\thetafn}[1]{\  \! \theta \left(#1\right)}
\newcommand{\tin}{\int_{-\infty}^{+\infty}\! \! \!\!\!\!\!}
\newcommand{\Tr}[1]{\mbox{Tr}\left(#1\right)}
\newcommand{\kb}{k_B}
\newcommand{\rad}{\mbox{ rad}}
\preprint{APS/123-QED}

\title{Emergence of local magnetic moments in doped graphene-related materials}
\author{P. Venezuela$^{(a)}$, R. B. Muniz$^{(a)}$, A. T. Costa$^{(a)}$, D. M. Edwards$^{(b)}$, S. R. Power$^{(c)}$ and M. S. Ferreira$^{(c)}$}
\email{ferreirm@tcd.ie}
\affiliation{
(a) Instituto de Fisica, Universidade Federal Fluminense, 24210-346 Niter\'oi - RJ, Brazil \\
(b) Department of Mathematics, Imperial College, London SW7 2BZ, UK \\
(c) School of Physics, Trinity College Dublin, Dublin 2, Ireland  }

\date{\today}

\begin{abstract}
Motivated by recent studies reporting the formation of localized magnetic moments in doped graphene, we investigate the energetic cost for spin polarizing isolated impurities embedded in this material. When a well-known criterion for the formation of local magnetic moments in metals is applied to graphene we are able to predict the existence of magnetic moments in cases that are in clear contrast to previously reported Density Functional Theory (DFT) results. When generalized to periodically repeated impurities, a geometry so commonly used in most DFT-calculations, this criterion shows that the energy balance involved in such calculations contains unavoidable contributions from the long-ranged pairwise magnetic interactions between all impurities. This proves the fundamental inadequacy of the DFT-assumption of independent unit cells in the case of magnetically doped low-dimensional graphene-based materials. We show that this can be circumvented if more than one impurity per unit cell is considered, in which case  the DFT results agree perfectly well with the criterion-based predictions for the onset of localized magnetic moments in graphene. Furthermore, the existence of such a criterion determining whether or not a magnetic moment is likely to arise within graphene will be instrumental for predicting the ideal materials for future carbon-based spintronic applications. 

\end{abstract}

                 
\maketitle
\bibliographystyle{abbrv} 

The controlled introduction of impurities into an otherwise pristine solid is one effective way of tailoring the electronic properties of a material. In systems such as thin films, nanowires and nanoparticles the effect of added impurities is expected to be particularly pronounced due to the reduced dimensionality of the host materials. From a theoretical point of view, predictions of how the physical properties of a system are affected by dopants are currently made based on state-of-the-art density functional theory calculations (DFT). These calculations often consider single impurities added to a unit cell with periodic boundary conditions, under the assumption that cells are sufficiently large and that impurities are not able to interact with their neighboring counterparts. There are, however, interactions that tend to become more long-ranged as the dimensionality is reduced, one of which is the interaction between magnetic impurities embedded in a metallic system. This raises the question of whether representing independent particles by single impurity unit cells is a valid assumption in the case of magnetic dopants in low-dimensional metallic structures.                                                       

One example in which this assumption is frequently made is in the case of magnetic objects in graphene-related materials, of growing interest lately due to potential use in spintronic applications.  As a matter of fact, a recent study has comprehensively investigated the magnetic properties of transition-metal atoms embedded in a graphene sheet \cite{arkady09} indicating complex magnetic behavior as one moves across the periodic table. One remarkable finding in this pioneering survey is the absence of a magnetic moment when one Fe atom substitutionally replaces a single carbon atom of the graphene sheet. Of all transition metal atoms, it is somewhat surprising that such an iconic magnetic element like Fe seems unable to develop a magnetic moment when immersed in graphene. Because the aforementioned DFT assumption of independent unit cells is used in the referred survey as well as in other studies of magnetic dopants in carbon-based structures \cite{portal1, portal2, ciraci-survey}, it is instructive to ask whether the intrinsic long-ranged interaction that arises between magnetic moments in low-dimensional metals might be responsible for interfering with some of the results recently reported. If so, this interference may spuriously suppress the formation of magnetic moments where they should actually exist. 

Lieb's theorem \cite{lieb} is often quoted to explain the magnetic properties of graphene. It shows that a net magnetization arises when there is an imbalance between the two non-equivalent sub-lattices composing the bipartite lattice of graphene. While this is a perfectly sound explanation for graphene flakes and ribbons, as well as for graphene sheets containing vacancies \cite{lopez}, it is not directly applicable to substitutionally doped impurities since Lieb's theorem assumes a homogeneous electron-electron interaction throughout the system. In particular, the considerably narrower $d$-band associated with transition-metal impurities makes the electronic interaction highly non-homogeneous, and another explanation for the origin of magnetic moments in doped graphene-related materials is required. The formation of a single local moment in a non-magnetic system has been generally addressed by several authors, and a criterion for its existence has been previously derived in different contexts \cite{heine}. Here we sketch an alternative derivation of this criterion, and generalize it to a pair of impurities in order to clarify the role played by the long range interaction between magnetic moments in low-dimensional systems. Furthermore, we show how this may affect first-principles calculations which artificially assume, for computational purposes only, that the system is translationally invariant. Although our focus is on doped graphene sheets, our conclusions result essentially from the hexagonal symmetry of the underlying lattice and are valid to graphene ribbons and flakes as well as nanotubes. 

We start by considering a single transition metal atom embedded in a non-magnetic host, which in our case is a pristine hexagonal lattice. We describe the electronic structure of the system by a Hubbard-like Hamiltonian $\hat{H} = H_0 + H_{int}$, where $H_0 = \sum_{ij\mu\nu \sigma} \gamma_{ij}^{\mu\nu} \, {\hat c}_{i\mu\sigma}^\dag \, {\hat c}_{j\nu\sigma}$ represents the electronic kinetic energy plus a spin-independent local potential, and $H_{int}$ is the electron-electron interaction term. The operator $\hat{c}_{i\mu\sigma}^{\dag}$ creates an electron with spin $\sigma$ in atomic orbital $\mu$ on site $i$. We assume that $H_{int}$ is an on-site interaction which takes place between electrons occupying the $d$ orbitals of the transition metal impurity only, and is neglected elsewhere. In this case, the matrix elements of the spin-dependent part of the one-electron hamiltonian reduces to $v_{\mu\nu}^{\sigma}=-\frac{1}{2} \Delta_{\mu} \delta_{\mu \nu} \sigma$, where $\Delta_{\mu}$ represents the local exchange splitting associated with orbital $\mu$, and $\sigma = \pm 1$ for $\uparrow$ and $\downarrow$ spin, respectively. For simplicity we shall assume that the on-site effective exchange integrals $U$ are the same for all $d$ orbitals, whence it follows that $\Delta_{\mu}$ is $\mu$-independent and equal to $\Delta = U M$, where $M$ is the local magnetic moment. 

To derive a local moment criterion we examine the stability of the non-magnetic state described by $H_0$ when $H_{int}$ is activated considering relatively small values of the exchange splitting. The energy cost involved in the formation of a local magnetic moment at the impurity site is given by 
 \begin{widetext}
 \begin{equation}
\Delta {\cal E}_1 =  \, {1 \over \pi} \, {\rm Im} \, {\rm Tr} \, \int_{- \infty}^{E_F} dE \, \left \{ -U \, {\cal G}_{0,0}(E) {\Delta \over 2}\left [1 - {\cal G}_{0,0}(E) \, {\Delta \over 2}\right ]^{-1} \, + \, {\rm log} \left [ \left( 1 + {\cal G}_{0,0}(E) \, {\Delta \over 2} \right) \, \left( 1 - {\cal G}_{0,0}(E) \,  {\Delta \over 2} \right)\right] \, \right \} \,\,,
\label{DeltaE1}
\end{equation}
\end{widetext}
where ${\cal G}_{0,0}(E)$ is the single-particle Green function for an electron with energy $E$ at the impurity site and $E_F$ is the Fermi energy. The trace operator is over the orbital degrees of freedom which in this case can be limited to the five $d$-orbitals. The first term of the integrand in Eq.(\ref{DeltaE1}) represents the reduction of the effective electron-electron interaction due to the appearance of a local spin imbalance at the impurity site, and the second term accounts for the increase of the corresponding electronic kinetic energy. The sign of $\Delta {\cal E}_1$ determines whether or not the non-magnetic state is unstable to a local magnetic moment formation. To derive a criterion for such an instability it is sufficient to expand Eq.(\ref{DeltaE1}) in powers of $\Delta$ to lowest order, which simplifies to $\Delta {\cal E}_1 = \{ - U \, \ell^2(E_F) \,  + \, \ell(E_F) \} \, (\Delta / 2)^2$, where $\ell(E_F) = {1 \over \pi} \, \int_{- \infty}^{E_F} dE \, {\rm Im} \, {\rm Tr} \, [{\cal G}_{0,0}(E)]^2$ is the local susceptibility. The formation of a local magnetic moment at the impurity is then energetically favorable when $\Delta {\cal E}_1 <0$, {\it i.e.}, when
\begin{equation}
\ell(E_F) > {1 \over U} \,\,.
\label{gsc1}
\end{equation}
This inequality sets the condition for the spontaneous formation of a single localized magnetic moment in a non-magnetic host. Written in terms of single-particle Green functions, it is model-independent and can be evaluated once the Hamiltonian is fully specified. The same criterion has been previously derived in other circumstances \cite{heine}, but here we have obtained it by total energy balance considerations because it provides an easier way to analyze the effect of more impurities. 

Let us now imagine that a second transition-metal impurity is added at site $m$. Similar steps may be taken to derive the following expression for the energy cost involved in the formation of local magnetic moments in the two impurities 
\begin{equation}
\Delta {\cal E}_2 = 2 \, \Delta {\cal E}_1 + {\Delta^2 \over 4 \pi} \, \int_{- \infty}^{E_F} dE \, {\rm Im} \, {\rm Tr} \, \left[{\cal G}_{0,m}(E) \, {\cal G}_{m,0}(E)\right] \,\,.
\label{DeltaE2}
\end{equation}
The energy cost $\Delta {\cal E}_2$ is not twice as large as $\Delta {\cal E}_1$ due to the interference between the two impurities. This is evident in the second term of Eq.(\ref{DeltaE2}), which contains Green function propagators between sites $0$ and $m$. Most remarkably, this interference term that arises naturally when we evaluate the energy cost for the formation of two separate magnetic moments is precisely the same as the expression for the RKKY coupling between magnetic impurities embedded in a metallic system \cite{prb1994, coupling1}. Bearing in mind that this coupling is negative (positive) when the magnetic moments are parallel (antiparallel) and that it decays very slowly with the impurities separation in the case of low dimensional systems, this additional interference term may have striking consequences to the criterion presented above. 

Consider for instance a hypothetical impurity that meets the inequality of Eq.(\ref{gsc1}), that is, an impurity that possesses a magnetic moment when immersed in the graphene lattice. Suppose that we add a second impurity of the same hypothetical element with the imposed constraint that both moments must be parallel to each other. The energy cost that was negative for a single impurity may become positive if the RKKY coupling favors an antiparallel alignment between the magnetic moments. In this case, $\Delta {\cal E}_2$ may become positive even though $\Delta {\cal E}_1 < 0$. If this occurs, the two magnetic impurities whose moments are forced to remain parallel may adopt an altogether non-magnetic configuration rather than the most favorable antiparallel alignment. This is a clear indication that the artificial imposition of parallel alignment may introduce spurious effects as far as the determination of the true ground state configuration is concerned.  

DFT-based calculations that consider one single magnetic impurity per periodically repeated unit cell implicitly impose that their magnetic moments, should they exist, must be mutually parallel. Because of the periodic boundary conditions, the energy cost (per impurity) $\Delta {\cal E}_N/N$ for inducing the spin splitting of $N$ equally spaced impurities becomes  
\begin{equation}
{\Delta {\cal E}_N \over N}= \Delta {\cal E}_1 + {\Delta^2  \over 4 N \, \pi} \,\sum_j^N \int_{- \infty}^{E_F} dE \, {\rm Im} \, {\rm Tr} \, \left [{\cal G}_{0,mj}(E) \, {\cal G}_{mj,0}(E)\right ] \,\,.
\label{DeltaEM}
\end{equation}
In this case the correction to the single-impurity contribution $\Delta {\cal E}_1$, which once again is assumed to be negative, is a sum of terms proportional to the pairwise RKKY interactions that may be positive and sufficiently large to reverse the sign of $\Delta {\cal E}_N/N$. One could argue that the RKKY interaction, being traditionally oscillatory as a function of separation, will alternate between negative and positive terms in the sum that appears in Eq.(\ref{DeltaEM}), which will then average out and never be able to reverse the sign imposed by $\Delta {\cal E}_1$. While this may be true in general, for graphene-based materials the underlying hexagonal atomic structure introduces a peculiar feature in the RKKY-like coupling that will seldom vanish the sum of  Eq.(\ref{DeltaEM}). It has been shown that the magnetic coupling between impurities embedded in graphene-related materials obeys the following rule \cite{carbon09, portal2}: $J_{A,A}/|J_{A,A}|=J_{B,B}/|J_{B,B}|=-J_{A,B}/|J_{A,B}|$, where $J_{A,A}$ ($J_{B,B}$) represents the Heisenberg-like magnetic coupling between impurities occupying the $A$ ($B$) sub-lattice sites of the hexagonal lattice. In other words, in the case of graphene-related materials the sign of the coupling between impurities occupying equivalent sites is the same regardless of their separation. Therefore, if the magnetic coupling between like sites favors the anti-parallel alignment between the magnetic moments, all terms appearing in the summation of  Eq.(\ref{DeltaEM}) will be positive, which may converge to a sufficiently large value capable of overturning the satisfied criterion for single impurities. In addition, in the case of nanotubes, we have shown that the coupling magnitude tends to decay rather slowly as $1/D$, where $D$ is the separation between magnetic impurities \cite{coupling1}. Such a slow decaying rate will turn the summation of  Eq.(\ref{DeltaEM}) into a non-convergent series in the limit $N \rightarrow \infty$, meaning that the correction will always surpass the magnitude of $\Delta {\cal E}_1$. 

The striking implications of this mathematical analysis are that spurious nonmagnetic solutions may be obtained if existing magnetic moments are artificially constrained to adopt a parallel alignment when they would spontaneously prefer to be antiparallel. This raises the question whether the recently reported absence of magnetic moments for Fe in graphene could be one such case \cite{arkady09}. It is worth stressing that more than correcting an inaccuracy, this would be a convincing indication of the inadequacy of the assumption, commonly used in DFT-calculations, of independent unit cells when dealing with magnetic dopants in carbon-based structures. One simple way of testing if the moment suppression is the result of the artificial constraint imposed by the periodic boundary conditions of the DFT scheme is to include more than one magnetic impurity per unit cell and allow them to adopt both parallel and anti-parallel alignments. In fact, in what follows we present DFT results for calculations comprising two impurities per unit cell and compare those with the results for a single impurity. 

Our DFT-calculations have been been made with the generalized gradient approximation \cite{pbe} for the exchange-correlation term. Troullier-Martins pseudo-potentials \cite{tm} and double-zeta polarization atomic orbitals \cite{siesta} have been used. The calculations were made with periodic boudary conditions and supercells comprising 4x4 and 7x7 graphene primitive cells. In these cases, one carbon atom of the graphene lattice was substituted by a single Fe atom impurity. The distance between an impurity and its image in the adjacent unit cell is 9.98 and 17.47 ${\rm \AA}$ for the 4x4 and 7x7 supercells, respectively. Unsurprisingly, our results are very similar to those previously reported \cite{arkady09}. We find that, when relaxed, the metal impurities are displaced outwards from the graphene surface by $1.14 \, {\rm \AA}$ and that no magnetic moment is observed. 

The results are completely different, however, when two impurities per unit cell are considered. To maintain the same impurity separation as before, we duplicate the unit cells along one direction (4x8 and 7x14 primitive cells). With two Fe impurities per unit cell, we have the freedom to start these calculations with magnetic moments in the anti-parallel configuration. Such an antiferromagnetic alignment between the Fe moments  is stable and energetically favorable, by 0.03 eV, when compared to the non-magnetic solution, which can be obtained by relaxing the spin-polarization. Remarkably, the substitutional Fe impurity does have a magnetic moment that is as large as 0.99 $\mu_B$. It is also instructive to calculate the energetics of the system in the configuration in which the Fe moments are parallel \cite{fefm}. In this ferromagnetic case the total energy is considerably higher than the antiferromagnetic configuration. Table I shows the total energy values obtained for Fe as well as Mn impurities in the ferromagnetic (FM), spin-unpolarized (SU) and antiferromagnetic (AF) configurations. For Fe, the total energies in descending order are (FM, SU, AF). Because the AF configuration is impossible to obtain with the single-impurity unit cell, the system adopts the next possible configuration, which shows no spin polarization. Alternatively, one can understand this in terms of Eq.(\ref{DeltaEM}), which means that the magnetic-coupling correction that arises due to the periodic boundary conditions is able to revert the sign imposed by $\Delta {\cal E}_1$ leading to a spurious suppression of the existing Fe magnetic moments. 

\begin{table}[htdp]
\begin{center}
\begin{tabular}{|c|c|c|}
 & Fe & Mn \\
FM & 0.14 &  0.04 \\
SU &  0.03 &  1.80 \\
\end{tabular}
\end{center}
\label{tabela}
\caption{Total energies, in eV, of the FM and SU configurations, for Fe and Mn impurities in graphene. All quantities are expressed relatively to the total energy of the AF configuration, which is the most energetically favorable for both impurities.}
\end{table}%

Also shown in Table I are the values for Mn impurities, for which the total energies in descending order are (SU, FM, AF). Another interesting result, not considered in reference \cite{arkady09}, is that the AF is the most energetically favorable configuration. Once again, this is easily understood by the single-impurity unit cell constraint that is unable to account for the AF alignment of the magnetic moments. In this case, the next possible configuration is the FM alignment, which again explains the results of reference \cite{arkady09}. In terms of Eq.(\ref{DeltaEM}), the magnetic-coupling correction for Mn impurities are not sufficient to overturn the sign determined by $\Delta {\cal E}_1$, which means that  $\Delta {\cal E}_N/N$ is still negative justifying the splitting of the spin-polarized bands into a FM configuration. 

This is an unmistakable proof of the potential problems that may arise when dealing with magnetic impurities in graphene-based structures through the standard DFT scheme of single-impurity unit cells. Furthermore, it is also a clear indication of the relevance of the coupling between magnetic impurities across graphene-related materials \cite{coupling1, carbon09, non-heisenberg, suppression, suppression2, rkky-outros, rkky-sarma, dynamic-coupling}. When this coupling is positive and sufficiently large to reverse the sign imposed by $\Delta {\cal E}_1$, the artificial constraints imposed by the periodic boundary conditions spuriously suppresses the magnetic moment that would spontaneously exist in isolation. In DFT-calculations of doped graphene-related materials, it is therefore of paramount importance to consider more than a single impurity per unit cell and study the energetics of all possible configurations, namely, FM, SU and AF. 

Finally, regarding the condition for the formation of a localized magnetic moment expressed by the inequality of Eq.(\ref{gsc1}), we can test its predictive power by applying it to the cases considered here. Written in terms of single-particle Green functions, the susceptibility $\ell(E_F)$ can be further simplified in the case of small spin-splittings ($\Delta \ll 1$) to $\ell(E_F) \approx \rho_0(E_F)$, where $\rho_0(E_F)$ is the spin-unpolarized local density of states (LDOS) at the impurity site evaluated at the Fermi level $E_F$. The value of $U \approx 1 {\rm eV}$, being primarily an atomic property, is fairly constant for all transition metal elements \cite{himpsel}. Therefore, we can use the LDOS obtained at our SU calculations and test whether the inequality of Eq.(\ref{gsc1}) is satisfied. Reassuringly, $U \rho_0(E_F) > 1$ for both Fe and Mn, indicating that both elements favor the formation of a magnetic moment when embedded within graphene. Further tests were carried out with Ni impurities. In this case, the low value found for $\rho_0(E_F)$ does not meet our criterion, suggesting that Ni atoms within graphene will not develop a magnetic moment. In fact, this is what we have found in our DFT-calculations with two Ni impurities per unit cell, which also agrees with previously reported results \cite{arkady09, portal1}. Such a good agreement with the predictions based on Eq.(\ref{gsc1}) indicates that SU calculations, which are considerably less time consuming than spin-polarized ones, can be carried out first to test whether a localized magnetic moment is likely to arise. If so, further spin-polarized calculations are required in which all possible configurations must be then considered. 

In summary, we have shown that the use of single-impurity-doped unit cells in DFT-based calculations is highly inappropriate to describe magnetically doped graphene and that it may lead to fundamentally erroneous results. This is a consequence of the inherently long range nature of the magnetic interaction between impurities, which makes the hypothesis of independent unit cells in such systems invalid. Graphene being a material of increasing popularity and DFT-calculations being the most widely used tool for studying the physical properties of materials, it is of paramount importance to account for this interaction when describing magnetically doped graphene and related structures. Furthermore, we present a mathematically transparent criterion for the formation of magnetic moments in graphene, something that has only been previously attempted on an {\it ad-hoc} basis \cite{arkady09}. The existence of a simple criterion that can tell whether or not a magnetic moment will arise when impurities are introduced to graphene-related materials is essential to predict which of these structures may be useful for spintronic applications.Ê

\end{document}